%% file: 3He_spectro_paper_draft.tex
\documentclass[%
 reprint,
 amsmath,amssymb,
 aps,
]{revtex4-2}
\pdfoutput=1
\newcommand{\beginsupplement}{%
        \setcounter{table}{0}
        \renewcommand{\thetable}{S\arabic{table}}%
        \setcounter{figure}{0}
        \renewcommand{\thefigure}{S\arabic{figure}}%
     }

\usepackage{graphicx}
\usepackage{transparent}
\usepackage{dcolumn}
\usepackage{bm}
\usepackage{color}
\usepackage{natbib}
\newcommand{\comment}[1]{}
\makeatletter
\def\maketitle{
\@author@finish
\title@column\titleblock@produce
\suppressfloats[t]}
\makeatother

\begin{document}

\preprint{APS/123-QED}

\title{The alpha and helion particle charge radius difference from spectroscopy of quantum-degenerate helium}

\author{Y. van der Werf}
\author{K. Steinebach}
\author{R. Jannin}
\author{H.L. Bethlem}
\author{K.S.E. Eikema}
\affiliation{LaserLaB, Vrije Universiteit Amsterdam.}

\date{\today}

\begin{abstract}
Accurate spectroscopic measurements of calculable systems provide a powerful method for testing the Standard Model and extracting fundamental constants. Recently, spectroscopic measurements of finite nuclear size effects in normal and muonic hydrogen resulted in unexpectedly large adjustments of the proton charge radius and the Rydberg constant. We measured the $2^3\mathrm{S}\rightarrow2^1\mathrm{S}$ transition frequency in a Fermi gas of $^3$He with an order of magnitude higher accuracy than before. Together with a previous measurement in a $^4$He Bose-Einstein condensate, a squared charge radius difference $r^2_h - r^2_{\alpha} = 1.0757(15)\ \mathrm{fm^2}$ is determined between the helion and alpha particle. This measurement provides a benchmark with unprecedented accuracy for nuclear structure calculations. A deviation of 3.6$\sigma$ is found with a determination~\cite{Schuhmann2023} based on spectroscopy of muonic helium ions.

\end{abstract}

\maketitle


\section{\label{sec:intro} Introduction}

Precise measurements of the energy level structure of few-electron atoms and molecules can be directly compared with \textit{ab initio} calculations including quantum electrodynamics (QED) and nuclear size effects. By comparing measurements on several transitions in different systems, one can determine the fundamental constants that are required for the calculations, and at the same time check the consistency of the Standard Model. This is a powerful method to search for yet unknown physics.

A cornerstone of such comparisons has long been atomic hydrogen, where the 1s-2s transition, measured so far with a $4\times10^{-15}$ relative accuracy \cite{Parthey2011}, serves as an accurate reference for level structure calculations.
Together with other precise transition measurements in the H atom, this is used to determine the Rydberg constant and the proton charge radius \cite{Mohr2016}. It was shown that the charge radius of the proton could be determined much more accurately \cite{Pohl2010} with spectroscopic measurements of the Lamb shift in muonic hydrogen (the bound state of a proton and a negative muon), than based on normal atomic hydrogen spectroscopy alone, or by electron scattering \cite{Gao2021,Yan2018}. This led to a surprising 4\% smaller recommended value for the proton charge radius and a $6\sigma$ adjustment of the Rydberg constant in CODATA~2018 ~\cite{Antognini2013,Tiesinga2021}. The new values were later confirmed with improved measurements on normal atomic hydrogen \cite{Beyer2017,Grinin2020,Bezginov2019}, although some disagreement still persists \cite{Fleurbaey2018,Brandt2022}.

It is interesting to look at the next simplest atom in the periodic table, helium. Based on spectroscopy and QED calculations~\cite{Patkos2021} of the two readily available isotopes, $^3$He and $^4$He, one can in principle determine the charge radii of the helion and alpha particle, respectively. However, even though state-of-the-art QED calculations reach the 7th order in the fine-structure constant for helium~\cite{Patkos2021}, complications from the two interacting electrons limit the accuracy to a level that makes it difficult to determine the nuclear charge radius from spectroscopy. The solution is to investigate the isotope shift of a transition in $^3$He and $^4$He instead. Most of the terms related to the electron interactions are nearly equal in the theory for both isotopes, enabling a sub-kHz accuracy for isotope shift  calculations~\cite{Pachucki2015,Pachucki2017}. When compared with the experimentally measured isotope shift, this enables an accurate determination of the squared nuclear charge radius difference between the helion and alpha particle.

In this work, we present a determination of the helion-alpha charge radius difference ($r^2_h - r^2_{\alpha}$) with unprecedented accuracy, based on a measurement of the doubly-forbidden, ultra-narrow, $2^3\mathrm{S}\rightarrow2^1\mathrm{S}$ transition. We perform our measurement in a quantum-degenerate Fermi gas of $^3$He atoms in the metastable $2^3\mathrm{S}_1\ (F=3/2)$ state, confined in an optical dipole trap (ODT). The ODT operates at the so-called `magic wavelength', where the two states involved in the transition experience the exact same energy shift.
By combining this measurement with our previous measurement of the same transition in a $^4$He Bose-Einstein condensate~\cite{Rengelink2018}, we obtain the most accurate determination of the squared charge radius difference between the alpha and helion particles.
Our measurement enables an accurate comparison of this difference with the one that can be obtained with spectroscopy of muonic He$^+$ ions ($\mu$He$^+$) ~\cite{Krauth2021,Schuhmann2023}.
Such a comparison provides a test of our understanding of both nuclear and atomic physics probed with electrons and muons, which is relevant in view of the muon g-2 anomaly \cite{Abi2021,Aoyama2020,Borsanyi2021}.

\section{Principle and experimental setup }
\begin{figure}[h]
\includegraphics[width=.49\textwidth]{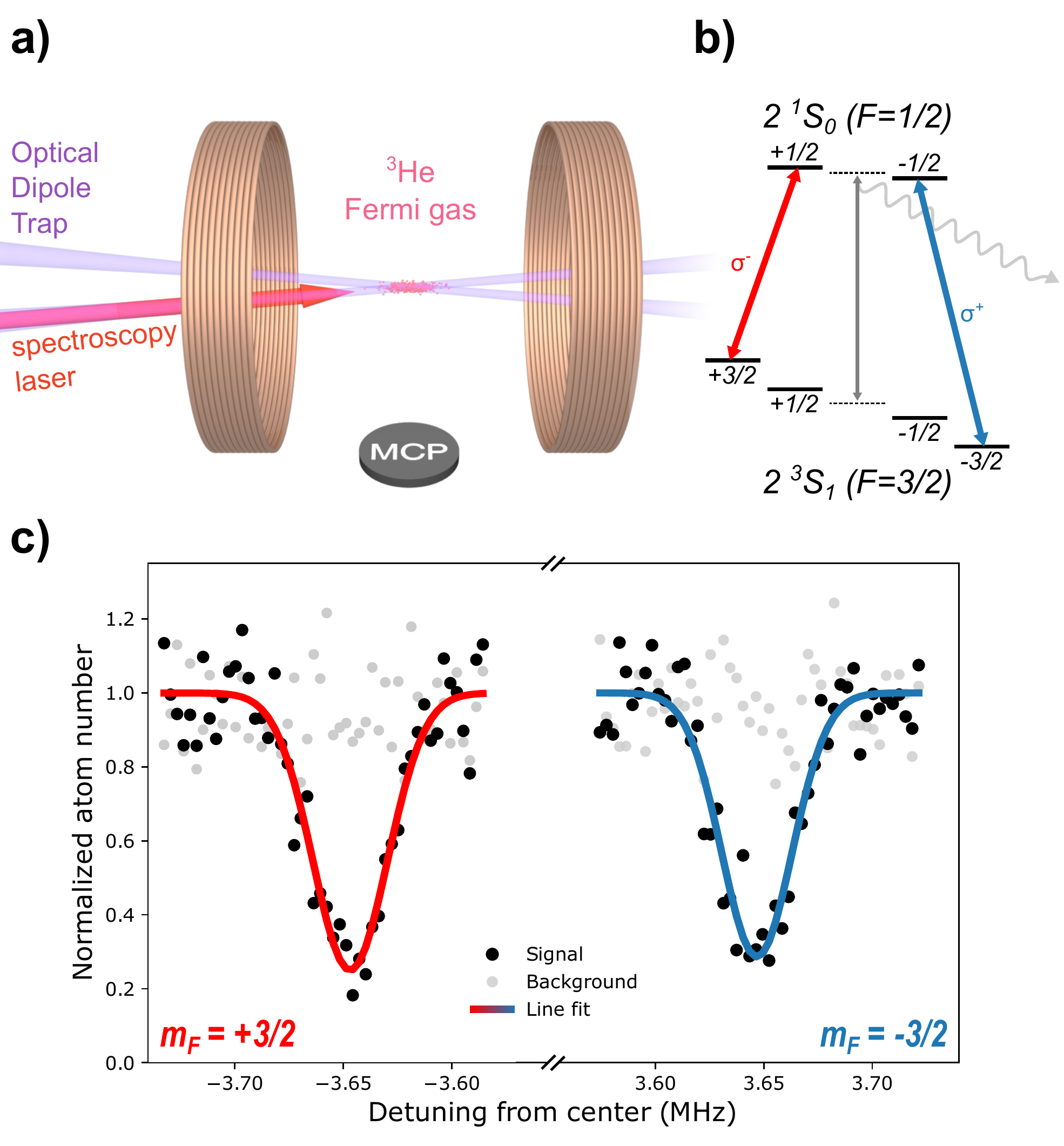}
\caption{\label{fig:typscan} \textbf{a)} Key elements of the experimental setup: a degenerate Fermi gas of $^3$He is loaded into the $319.8$ nm optical dipole trap and exposed to the spectroscopy laser, after which the remaining atoms are detected using the micro channel plate (MCP) detector. The quantization axis is defined and maintained by a homogeneous magnetic field generated by a pair of Helmholtz coils.
\textbf{b)} We alternate between measuring the transitions from the two spin-stretched magnetic field projections, to cancel the Zeeman shift due to the magnetic field.
\textbf{c)} Typical example of a recorded spectrum measured as a loss of $2^3\mathrm{S}_1$ atoms from the trap, after illumination for 3 seconds with the excitation laser for each data point. The width of the transition is dominated by the Doppler effect due to fermionic quantum statistics \cite{Jannin2022}. The Zeeman-free transition, represented by the vertical gray arrow in \textbf{b}, is found by taking the average of the two measured lines. Other systematic shifts are accounted for by measuring spectra under different conditions.}
\end{figure}

The spectroscopic measurements in $^3$He are performed with an experimental apparatus that is similar to that used in earlier work \cite{Rengelink2018}, and details of it are given in the Supplementary Material. In short, we prepare degenerate Fermi gases of $^3$He in a magnetic trap by sympathetic cooling with laser cooled $^4$He, and load the $^3$He Fermi gas into an optical dipole trap close to the magic wavelength of $319.8$ nm, as illustrated in Figure \ref{fig:typscan}\textbf{a}. There the gas is exposed to $1557$ nm spectroscopy light for several seconds to induce transitions from the $2^3\mathrm{S}_1\ (F=3/2)$ to the $2^1\mathrm{S}_0\ (F=1/2)$ state, which decays to the untrapped ground state. The resulting loss of atoms is a measure of the level of excitation, which is detected by releasing the remaining atoms from the dipole trap so that they fall due to gravity on a 2-stage 14.5 mm diameter micro-channel plate detector. 
We repeat the cold atom preparation and detection sequence for the different laser frequencies, alternated with background measurements without spectroscopy light. In this manner a spectrum is built up as illustrated in Figure \ref{fig:typscan}\textbf{c} starting from the spin-stretched $m_F=\pm3/2$ states as shown in Figure \ref{fig:typscan}\textbf{b}. These two states have exactly opposite linear Zeeman shifts, which is used to cancel the systematic effect from the non-zero magnetic field (see Methods section for details).
The width of the line profile is dominated by Fermi-Dirac statistics, containing effects of Doppler broadening by the Fermi temperature and Pauli blockade in the excitation dynamics \cite{Jannin2022}. 
The required analysis is therefore fundamentally different from that encountered in our earlier work on a $^4$He Bose-Einstein condensate \cite{Rengelink2018}. 

\section{Measurements and data analysis}
\begin{figure}[ht]
    \includegraphics[width=.49\textwidth]{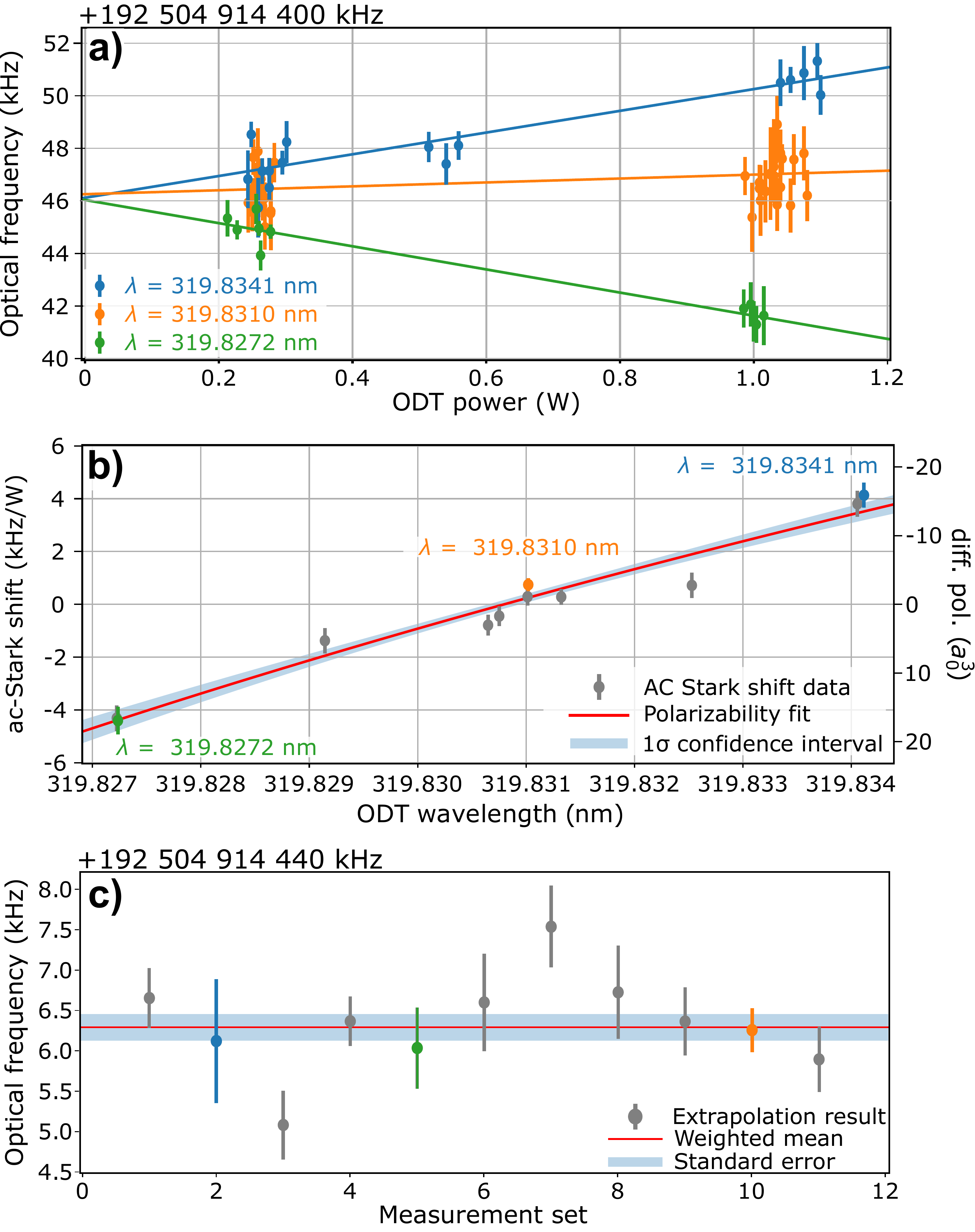}
    \caption{Overview of the measurement results. \textbf{a)} Determination of the differential ac-Stark shift on the transition frequency due to the trapping potential, shown for three different wavelengths of the ODT. The regression results for these three sets are indicated in \textbf{b} and \textbf{c} with corresponding colours. \textbf{b)} From the trap power regressions at different ODT wavelengths the magic wavelength, for which the power dependence of the transition frequency vanishes, can be found. The axis on the right shows the corresponding differential polarizability of the $2^1\mathrm{S}$ and $2^3\mathrm{S}$ states (See Supplementary Material). \textbf{c)} The obtained transition frequencies at zero trapping- and spectroscopy power resulting from the different sets, are averaged to give the final result in Table \ref{tab:errorbudget}, shown as the solid red line.}
    \label{fig:transition_result}
\end{figure}

The most important systematic effect on the measured transition frequency is due to the ac-Stark effect induced by the spectroscopy and optical trapping laser beams. We characterize this influence  by varying the power of both beams.
To determine the magic wavelength where the shift of the trapping laser is zero, we measure the variation of the ac-Stark shift of the $2^3\mathrm{S}_1 (F=3/2)\rightarrow2^1\mathrm{S}_0 (F=1/2)$ transition as a function of the trap power and wavelength. Figure \ref{fig:transition_result}\textbf{a} shows such a measurement for three different trap wavelengths. We fit the resulting power-dependent shifts with the polarizability curve calculated in \cite{Notermans2014}, as shown in Figure \ref{fig:transition_result}\textbf{b} (details are given in the Methods section). The zero-crossing, indicating the magic wavelength, occurs at $319.830\,80(15)$ nm, which is in good agreement with the value of $319.830\,2(7)$ nm from relativistic configuration-interaction calculations \cite{Wu2018}. The estimated size of non-scalar contributions to the polarizability are well below the level of the measurement accuracy (see Supplementary Material). The difference between the magic wavelength of $^3$He and $^4$He \cite{Rengelink2018} is $43.6(6)$ GHz, which also agrees very well with the predicted $43.1(7)$ GHz shift calculated earlier~\cite{Notermans2014}. 

For each set of data at a fixed ODT wavelength, the powers of both the trap and spectroscopy laser are varied and a multiple-regression analysis is performed to extrapolate the measured frequencies to zero overall laser power. The averaged value of all these sets gives the measured transition frequency, as is shown in Figure \ref{fig:transition_result}\textbf{c}. In order to obtain the unperturbed energy splitting between the $2^3\mathrm{S}_1 (F=3/2)$ and $2^1\mathrm{S}_0 (F=1/2)$ states from these measurements, several more systematic effects and corrections are taken into account, with their contributions summarized in Table \ref{tab:errorbudget}. An elaborate investigation of these systematics is included in the Supplementary Material.

\begin{table}[h]
\caption{\label{tab:errorbudget}%
Summary of the corrections to the measured $2^3\mathrm{S}_1 (F=3/2)\rightarrow2^1\mathrm{S}_0 (F=1/2)$ transition and error budget of the different systematics. Details on the different systematic effects are given in the Supplementary Material.
}
\begin{ruledtabular}
\begin{tabular}{l r l}
\textrm{Contribution}&
\multicolumn{1}{r}{\textrm{Value (kHz)}}&
\multicolumn{1}{l}{\textrm{Error (kHz)}}\\
\colrule
\textrm{Measured frequency} & $192\,504\,914\,446.291 $ &  0.165\\
\textrm{Recoil shift} & -27.276 & 0.0 \\
\textrm{Cs clock} & -0.055 & 0.01\\
\textrm{Blackbody Radiation} & $<0.005$ & \\
\textrm{Collisional Shift} & $<0.001$  & \\
\textrm{Quantum Interference} & $< 8\times 10^{-5}$ & \\ 
\textrm{Second-order Zeeman} & $< 5\times10^{-5}$ & \\ 
\textrm{dc-Stark shift} &  $< 1\times10^{-6}$ & \\ 
\textrm{Second-order Doppler} & $< 2\times10^{-9}$ & \\ 
\hline
\textrm{Total:} & $192\,504\,914\,418.96\,\  $ & 0.17\\
\textrm{van Rooij \textit{et al.} \cite{VanRooij2011a}:} & $192\,504\,914\,426.7\,\ \,\ $ & 1.5 \\
\end{tabular}
\end{ruledtabular}
\end{table}

\section{Experimental results}
We obtain a $2^3\mathrm{S}\rightarrow2^1\mathrm{S}$ transition frequency for $^3$He of $192\,504\,914\,418.96(17)$ kHz, which is roughly an order of magnitude more precise than our previous determination ~\cite{VanRooij2011a}, and constitutes the most accurate frequency measurement in helium to date. The new determination presents a $5\sigma$ discrepancy with the former value. Based on new insights and measurements, we can explain this as a result of the interplay between the thermodynamics of Fermi gases and the ac-Stark shift imposed by the trapping potential. In the previous measurement \cite{VanRooij2011a}, the dipole trap was at 1557 nm, far away from the magic wavelength. Therefore a large inhomogeneous differential ac-Stark shift was imposed on the atoms due to the finite extent of the Fermi gas over the intensity profile of the trap. This typically causes an asymmetric line profile that is well described by \cite{Juzeliunas2001}, with a `thermodynamic shift' from the actual transition frequency which depends on the parameters of the gas. However, at the time of the former measurements, this asymmetric profile could not be resolved within the laser bandwidth \cite{VanRooij2011a}, and the thermodynamic shift was therefore not correctly taken into account. 
With new measurements in a $1557$ nm dipole trap, now resolving the asymmetric line profiles, we verified that the model from \cite{Juzeliunas2001} correctly accounts for the thermodynamic shift, producing a measured frequency in agreement with the new result presented here. 
Using this knowledge, we have corrected the $^3$He measurements of \cite{VanRooij2011a} for the thermodynamic shift and re-performed the ac-Stark extrapolation, resulting in a very good agreement with the results obtained in this work with the magic wavelength trap. In the Supplementary Material a detailed description of this analysis is given for both the 1557 nm and magic wavelength dipole traps. For the magic wavelength trap used in this work the effect of the thermodynamic shift on the determined transition frequency is negligible.

\begin{figure*}[t]
    \centering
    \includegraphics[width=\textwidth]{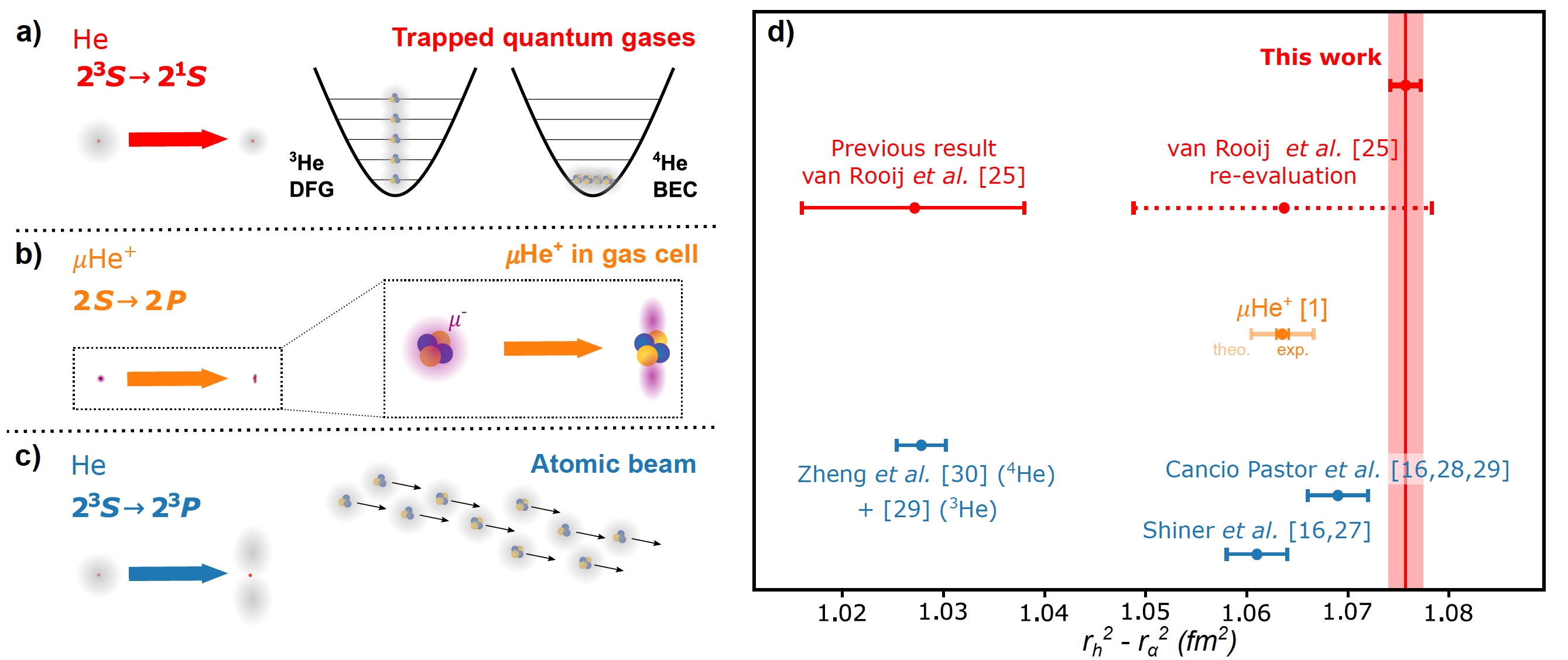}
    \caption{Determinations of the squared charge radius difference $r^2_h - r^2_{\alpha}$ through precision spectroscopy of helium and muonic-helium ions. The sensitivity to the nuclear sizes comes from the change in electron orbital between the two states involved in the measured transitions. \textbf{a)} Our work is based on the narrow $2^3\mathrm{S}\rightarrow2^1\mathrm{S}$ transition, measured in trapped quantum gases, which have distinctly different quantum statistical behaviour between the two isotopes. 
    \textbf{b)} In muonic helium ions, a strong enhancement of the finite size effect is used to directly extract the nuclear charge radii from the $2\mathrm{S}\rightarrow2\mathrm{P}$ Lamb shift.
    \textbf{c)} The determinations based on the $2^3\mathrm{S}\rightarrow2^3\mathrm{P}$ transition are performed in an atomic beam. A major challenge for this transition is the natural linewidth of $1.6$ MHz. 
    \textbf{d)} Comparison of $\delta r^2 = r^2_h - r^2_{\alpha}$ from the different experiments. An earlier large discrepancy between the value based on $2^3\mathrm{S}\rightarrow2^1\mathrm{S}$~\cite{VanRooij2011a} and the $2^3\mathrm{S}\rightarrow2^3\mathrm{P}$ ~\cite{Shiner1995,CancioPastor2012} isotope shift measurements is now resolved (see the main text and the re-evaluated measurement point). 
    However, our new improved result ("This work") deviates by 3.6\,$\sigma$ from that obtained with $\mu$He$^+$~\cite{Krauth2021,Schuhmann2023}}
    \label{fig:DNCR} 
\end{figure*}

The improved $^3$He transition frequency can be combined with the result obtained in a $^4$He Bose-Einstein condensate \cite{Rengelink2018} and current QED theory \cite{Pachucki2017} to obtain a more accurate squared nuclear charge radius difference $\delta r^2 = r^2(^3\mathrm{He}) - r^2(^4\mathrm{He}) = r^2_h - r^2_{\alpha} = 1.0757(15)\ \mathrm{fm^2}$ between the two helium isotopes. This is shown in Figure \ref{fig:DNCR}\textbf{c} together with other determinations of this quantity. 
We included also the re-evaluation of the measurement of~\cite{VanRooij2011a} based on the corrected $^3$He transition frequency as described earlier, which is in full agreement with the new result. It resolves a long-standing large disagreement between \cite{VanRooij2011a} and the measurements from \cite{Shiner1995} and \cite{CancioPastor2004,CancioPastor2012}.

There is a $2\sigma$ to $4\sigma$ discrepancy of our result with the two independent determinations involving the $2^3\mathrm{S}\rightarrow2^3\mathrm{P}$ transition \cite{Shiner1995,CancioPastor2012}. However, these measurements disagree with each other on a similar level. They require the same QED theory for the extraction of $\delta r^2$, so the disagreement originates most likely on the side of the experiment. It has been suggested~\cite{Marsman2015,Zheng2017} that quantum path interference effects might not have been correctly considered in~\cite{CancioPastor2012}. The $^4$He measurement from~\cite{Zheng2017} raises another question, as it deviates by $20\sigma$ from \cite{CancioPastor2004} and, in combination with the $^3$He result from~\cite{CancioPastor2012}, yields a value for $\delta r^2$ that is much smaller than any of the other experiments. Recently, a systematic effect was identified, due to a slit in the experiment of~\cite{Zheng2017}, that could explain the deviation~\cite{Hu2023}.

Our new determination of $\delta r^2$ enables an accurate comparison with the one obtained recently from $\mu$He$^+$ spectroscopy. With $\mu$He$^+$, the charge radius of both the alpha particle \cite{Krauth2021} and helion particle \cite{Schuhmann2023} have been determined independently with high accuracy, and they are in agreement with the much less accurate electron scattering data~\cite{Sick2015}. The resulting $\delta r^2$ from $\mu$He$^+$ spectroscopy is $\delta r^2 = 1.0636(6)(30)\,\mathrm{fm^2}$~\cite{Schuhmann2023}, with a combined experimental and theoretical uncertainty of $0.0031\,\mathrm{fm^2}$. When we compare this to our determination of $1.0757(15)\,\mathrm{fm^2}$ from normal helium, a deviation of 3.6$\sigma$ is observed. It would require an adjustment of 1.9 kHz on our measured $^3$He$-^4$He isotope shift to bring both determinations into agreement within 1~$\sigma$. This is much larger than our total uncertainty of 263 Hz for the isotope shift measurement. It should be noted that the errorbar of the muonic result is dominated by the theory used to determine the finite-size contribution to the measurement. The comparison with normal helium primarily constitutes a test of the different QED effects in both experiments and especially polarization of the nucleus that dominates the uncertainty for $\mu$He$^+$. 
Another possibility for the disagreement could be that there is a difference between muons and electrons besides their mass. 

\section{Conclusion and outlook}
Our measurement in a $^3$He Fermi gas presents the most accurate spectroscopy determination in helium and, together with the previous determination in $^4$He, results in the most accurate determination of $r^2_h - r^2_{\alpha}$. The determination of the magic wavelength in $^3$He provides an overall QED benchmark. 
It is now possible to do a high-precision comparison between the alpha and helion particle charge radii, based on spectroscopy of both electronic and muonic atoms.  The values for $r^2_h - r^2_{\alpha}$ based on He and $\mu$He$^+$ spectroscopy differ by 3.6~$\sigma$, for which there is currently no explanation. 
The presented nuclear charge radius comparison encompasses a wide range of physics, including Bose and Fermi statistics, bound-state QED theory of the electron and muon, and intricate QED effects like 2- and 3-photon-exchange and nuclear polarizability for $\mu$He$^+$. Therefore it is an extensive test of our current understanding of physics, including lepton universality, and it provides new input for developing more accurate nuclear structure models. 

\section{Acknowledgements}
We would like to thank the late Wim Vassen for inititating this research, and dedicate the paper to him. We would also like to thank NWO (Dutch Research Council) for funding through the Program 'The mysterious size of the proton' (16MYSTP) and Projectruimte grant 680-91-108. We thank Rob Kortekaas for technical support.

\newpage
\beginsupplement

\include{3He_spectro_Supplementary}

\bibliographystyle{apsrev4-1}
\bibliography{3He_spectro_bib}

\end{document}

%% file: 3He_spectro_Supplementary.tex
\title{Supplementary Material to: \\ Alpha and Helion Particle Size Difference Determination via Spectroscopy on Quantum-Degenerate Helium}

\maketitle

\section*{Methods}

\subsection{\label{methods:Sequence}Experimental Sequence}

To produce a cold sample of metastable He atoms, we start from a DC discharge source to populate atoms in the metastable $2^3\mathrm{S}_1$ state, producing a supersonic beam that is decelerated by a Zeeman slower and from which we load a magneto-optical trap (MOT). After a few seconds of MOT loading, the beam is turned off and a cloverleaf magnetic trap is switched on where first another laser cooling step is used to approach the Doppler cooling limit, before forced evaporative cooling is performed. With an RF sweep the most energetic atoms are transferred to an untrapped Zeeman substate, and the remaining atoms re-thermalize through collisions leaving a colder sample. 

The absence of $s$-wave collisions between identical fermions, due to the Pauli exclusion principle, makes direct evaporative cooling of spin-polarized $^3$He impossible. We therefore use a mixture with bosonic $^4$He to sympathetically cool the fermionic $^3$He component to quantum degeneracy. The $^4$He component is then removed from the trap leaving a pure Fermi gas of $^3$He, which we transfer to the optical dipole trap (ODT) by adiabatically turning off the magnetic trap field. In the ODT, with typical trap frequencies 
between $10$ to $30$ Hz and $0.15$ to $0.25$ kHz along the axial and radial directions, respectively, we are then left with a degenerate Fermi gas composed of $10^5$ to $10^6$ atoms at a temperature of $0.1$ to $0.6$ $\mu$K and a degeneracy $T/T_F$ ranging between $0.25$ and $0.6$. 
We then turn on a homogeneous bias magnetic field of a few Gauss to maintain a quantization axis and prevent depolarization of the sample, and after a 3 seconds spectroscopy step, the dipole trap is released and the expanding gas falls under gravity on the micro-channel plate detector (MCP). Here the metastable atoms release their $\sim 20$ eV internal energy, producing a time-of-flight trace from which the temperature, chemical potential, and number of atoms of the gas are extracted.
This sequence is followed by the same procedure without spectroscopy light to measure the fluctuations in the background number of atoms. This is repeated for different spectroscopy laser frequencies, switching between realisations where the atoms are kept in the $|F=3/2,m_F=+3/2\rangle$ state, or transferred to the $|F=3/2,m_F = -3/2\rangle$ state via adiabatic transfer with an RF sweep. In this way, spectra are recorded simultaneously for both spin states, based on the remaining number of atoms normalized to the background. We fit two Gaussian peaks, which is a good approximation of the spectral line shape for an ODT close to the magic wavelength \cite{Jannin2022}, and extract the center of the two lines. Such spectra are recorded for powers of the ODT ranging between $0.25$ and $1.1$ W, and the spectroscopy laser between $20$ and $90$ mW, in order to correct for the ac-Stark shifts.

\subsection{Data analysis and statistics}

Sets of measurements at varying laser powers were collected for different wavelengths of the optical dipole trap. For each set, the transition frequency is found by a weighted least-squares regression, which is used to extrapolate the data to zero laser power. From this regression also follows how strongly the ac-Stark shift depends on the trap power, which we use to find the magic wavelength. This power dependence is fitted as a function of trap wavelength using the polarizability curve calculated in \cite{Notermans2014}, with an offset on the wavelength and a coefficient converting polarizability into ac-Stark shift variation as the two fit parameters. The polarizability $\alpha$ is expressed in atomic units and can be converted to SI units by multiplication with $4\pi\epsilon_0 a_0^3$ ($\epsilon_0\approx8.854\times10^{-12}\ \mathrm{F/m}$ is the vacuum electrical permittivity and $a_0=0.0529$ nm the Bohr radius). We use a polarizability curve calculated for $^4$He, but as the main difference with $^3$He is the mass-dependent isotope shift of the levels, the overall shape is the same at the current level of accuracy and only the value of the magic wavelength is different. The value for the magic wavelength of $319.83080(15)$ nm results from the determined zero crossing of the fit function, with the error based on the $\chi^2$ of the fit to the data. From the conversion coefficient, we can relate the ac-Stark shift to the polarizability, and thereby find the average laser intensity (given $1$ W of power) responsible for the shift to be $5.5\times 10^7\ \mathrm{Wm^{-2}}$, which is slightly lower than the peak intensity of ${\sim}10^8\, \mathrm{Wm^{-2}}$ at the center of the dipole trap. This difference is mainly caused by the fact that the $^3$He Fermi gas is distributed over the full trap volume, and therefore extends into the region of lower light intensity, reducing the overall ac-Stark shift the gas experiences. This same effect caused the thermodynamic shift on the ac Stark extrapolation of \cite{VanRooij2011a}, but it does not affect the determination of the transition frequency in the magic wavelength trap, as is explained in detail later in this Supplementary Material.

For the determination of the final transition frequency, we take the weighted average value of all the individual extrapolated values, and take the standard error of the mean. To account for the variance of the data, this standard error is multiplied by the square-root of the reduced $\chi^2$, to obtain the final error. This error from the extrapolated values intrinsically contains both the contributions of the measurement statistics and uncertainty in the systematic shifts. To de-correlate these contributions, we have performed a single regression fit to all the data, including the trap and spectroscopy laser power, as well as the trap wavelength, as the independent variables. We then used the covariance matrix $C_{jk}$ of the regression coefficients $a_j, \cdots, a_k$ to calculate the confidence interval of the regression $y=f(\mathbf{x},a_j,\cdots,a_k)$, defined as:
\begin{equation}
    \sigma_y^2(\mathbf{x}) = \sum_{j} \sum_{k} C_{jk} \frac{\partial y}{\partial a_j}\Big|_\mathbf{x}\ \frac{\partial y}{\partial a_k}\Big|_\mathbf{x},
\end{equation}
where $\mathbf{x}$ is the vector of independent variables.
By minimizing $\sigma_y^2(\mathbf{x})$, we find that at a spectroscopy laser power of $\sim75$ mW, and an ODT with a power of $0.46$ W and wavelength of $319.83085$ nm, the uncertainty in the regression is purely dominated by statistics and is $43$ Hz.

The calculation of the differential squared nuclear charge radius, $\delta r^2$, is based on the method and calculated values from \cite{Pachucki2017}. We calculate the isotope shift between the $^3$He and $^4$He \cite{Rengelink2018} results, subtracting the hyperfine shift of the $2^3\mathrm{S}_1 (F=3/2)$ level \cite{Rosner1970}. We subtract the QED calculation of the isotope shift based on point nuclei \cite{Pachucki2017} and from the difference we find the finite nuclear size contribution to the shift. This contribution is then divided by a proportionality constant as evaluated in \cite{Pachucki2017}, expressing the sensitivity of the transition to the nuclear finite size effect, to find the resulting $\delta r^2$.

\subsection{\label{methods:Frequency}Frequency Metrology}

For the frequency stability and metrology, we use the transfer-lock to an ultrastable reference laser (Menlo Systems ORS1500, $\leq2$ Hz within $1$ s), via an optical frequency comb, which has been described in earlier works \cite{Rengelink2018}. We generate beatnotes of both the spectroscopy laser at $1557$ nm and the reference laser at $1542$ nm with their respective modes of the frequency comb. The beatnote of the spectroscopy laser is then downmixed with a direct digital synthesizer (DDS) and mixed with the reference laser beatnote, creating a 'virtual beatnote' at exactly $10$ MHz, which is phase-locked to the $10$ MHz reference from a Cs atomic clock. By tuning the downmixing frequency of the DDS, the frequency of the spectroscopy laser is changed to maintain the $10$ MHz virtual beatnote frequency, and thus we can build up our spectra. The resulting short-term linewidth of the spectroscopy laser is $5$ kHz and is mainly limited by an unstabilized fibre-link from the laser infrastructure laboratory to the metastable He experiment \cite{Rengelink2018,Notermans2016}.
The reference laser drifts in optical frequency on the long term, at an average rate of $\sim2$ kHz per 24 hours. This drift is characterised by monitoring the two beatnotes of the transfer lock, together with the frequency comb repetition rate and carrier-envelope-offset frequency, on a set of zero-deadtime counters (K+K Messtechnik). By a fit to the drifting beatnote counter data combined with the downmixing DDS frequencies, we extract the absolute optical frequency of the spectroscopy laser for every individual point of the recorded spectra, and fit the spectra in terms of this absolute optical frequency.
To correct for deviations of the Cs clock from the definition of the SI second, we compared the Cs clock and GPS pulse-per-second signals on an individual counter over the full measurement campaign. Based on this comparison we determined an offset of the Cs clock with respect to the GPS definition of the SI second at the level of $\sim2.9\times10^{-13}$, resulting in the correction of $55$ Hz shown in Table \ref{tab:errorbudget}. As a conservative error estimate for this correction, we have used the specified $5\times10^{-14}$ stability floor of the Cs clock. The frequency offset of the GPS satellites to the NIST frequency standard over the period of the measurements was $<1\times10^{-15}$ \cite{NIST2022}, so no further correction is needed.

\newpage
\section*{Supplementary Information}
The first part of this Supplementary Material presents an analysis of the ac-Stark shift induced by a non-magic wavelength dipole trap on a Fermi gas to explain the discrepancy between the result from the main text and the previous measurement \cite{VanRooij2011a}. With these new insights, the data from \cite{VanRooij2011a} are re-evaluated.
The second part gives estimations of several possible systematic influences on the measurement results, which are summarized in Table \ref{tab:errorbudget} of the main text.

\subsection*{\label{methods:lineshape}Spectral lineshape and ac-Stark shift}

\begin{figure}[ht]
    \centering
    \includegraphics[width=.49\textwidth]{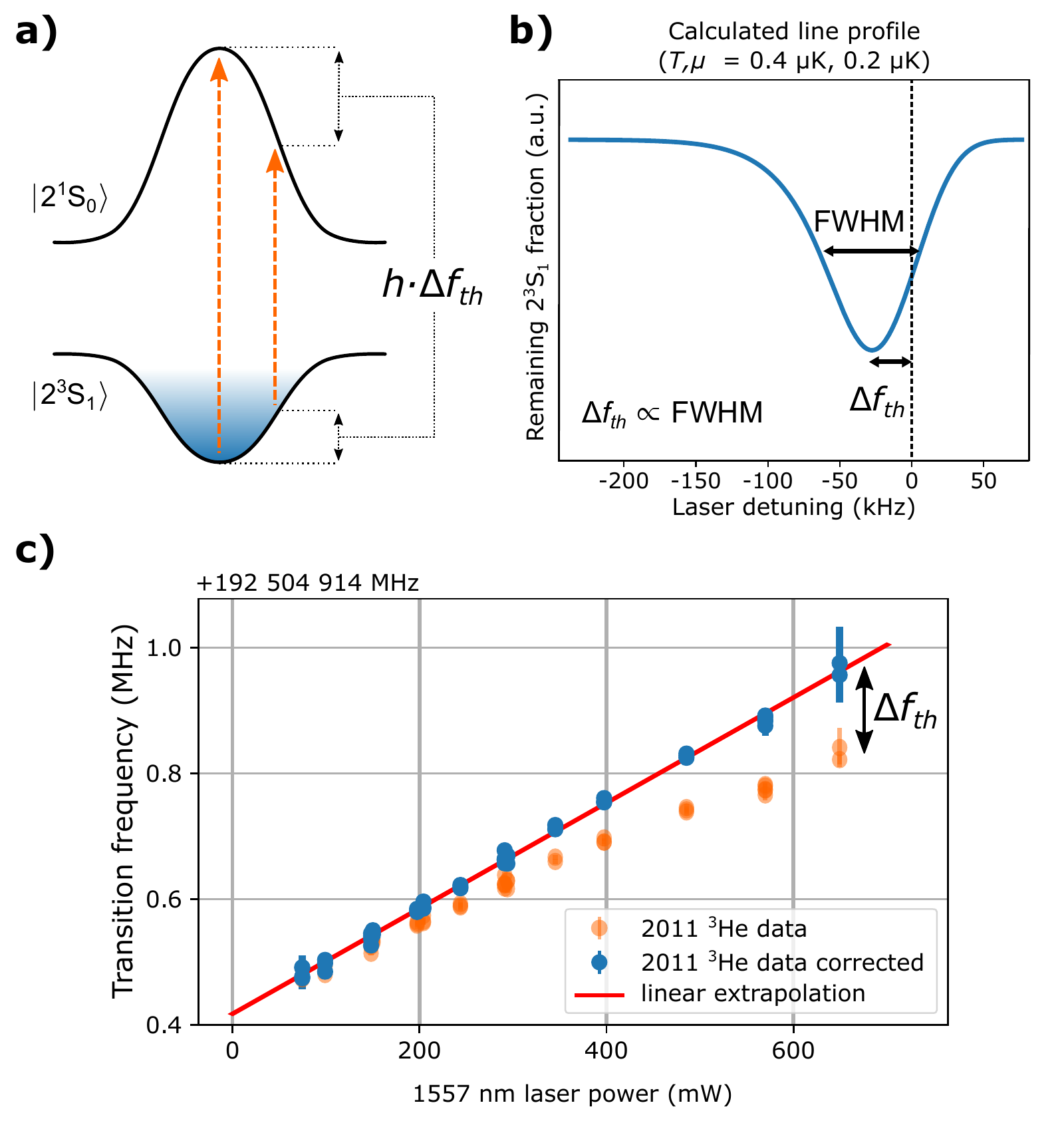}
    \caption{Thermodynamic contribution to the differential ac-Stark shift in a non-magic wavelength ODT. \textbf{a)} In a 1557 nm trap the two spectroscopy states experience opposite ac-Stark shifts. Due to the finite extent over the trapping region, the Fermi gas experiences an inhomogeneous trap light intensity, and the overall differential ac-Stark shift is lower than expected from the peak laser intensity and depends on the Fermi gas parameters. \textbf{b)} The spectral lineshape for this situation is well described by the model developed in \cite{Juzeliunas2001}, an example is shown for a temperature $T = 2\ \mathrm{\mu K}$ and chemical potential $\mu = 1\ \mathrm{\mu K}$. The precise line shape depends on the gas parameters, but the width and peak position show a strong correlation, independent on $T$ and $\mu$. \textbf{c)} Regression of the ac-Stark shift for the measurements from \cite{VanRooij2011a}. The orange points are the original data, showing a nonlinear dependence on the ODT power. With our current understanding, a record of the full width at half maximum (FWHM) of these measurements was used to calculate the thermodynamic contribution $\Delta f_{th}$ to the differential ac-Stark shift and correct the measurements (shown in blue). The linear extrapolation to zero power of the corrected data results in a shift-free transition frequency at $192\,504\,914\,417.6(2.0)$ kHz.}
    \label{fig:Thermo_shift}
\end{figure}

As mentioned in the main text, the main contribution to the discrepancy between our former result for the $^3$He $2^3\mathrm{S}_1\rightarrow2^1\mathrm{S}_0$ transition \cite{VanRooij2011a} and this work is due to the interplay between the ac-Stark shift induced by the trapping potential and the phase space distribution of the Fermi gas. The $1557$ nm optical dipole trap used in \cite{VanRooij2011a} caused a strong, inhomogeneous, differential ac-Stark shift on the transition, as illustrated in Figure \ref{fig:Thermo_shift}\textbf{a}, and the line profile for such a situation reflects both the momentum distribution of the gas through the Doppler effect, as well as the spatial distribution due to the variation of the light intensity along the trapping potential, which was later found \cite{Notermans2016} to be well described by \cite{Juzeliunas2001}. The typically asymmetric line profile resulting from this spatial distribution, of which an example is shown in Figure \ref{fig:Thermo_shift}\textbf{b}, could at the time not be resolved within the $90$ kHz bandwidth of the spectroscopy laser. The effect was therefore only observed through a shift of the centroid of the spectrum, appearing as a nonlinear dependence on the trap power, as can be seen in the original data in Figure \ref{fig:Thermo_shift}\textbf{c}. This nonlinear dependence has been investigated at the time, and was attributed to a depletion effect in the loading of the dipole trap, beyond a certain trap depth. The fit to the ac-Stark effect was then limited to trap depths below this alleged depletion limit, and a linear extrapolation was used. This argumentation is not consistent with the observed effect, as one would expect that a Fermi gas of the same number of atoms, temperature, and chemical potential inside a deeper trap would result in a larger overall ac-Stark shift. In contrast, it can be seen from the ac-Stark shift measurements that the deviation from linearity shows an overall reduction of the shift as the power increases, whereas an overall increase would be expected based upon the presented arguments.

For this work, we have re-investigated the nonlinear dependence of the ac-Stark shift on the trap power by performing new measurements in a $1557$ nm ODT, this time using the narrow excitation laser provided by the transfer lock described in Section \ref{methods:Frequency} of the Methods, and have resolved the asymmetric line profiles obeying the behaviour described by the model from \cite{Juzeliunas2001}. From the fit of the line profile with this model, the shift related to the peak intensity (corresponding to the center of the dipole trap) can be extracted. A linear extrapolation to zero power of these determinations then results in a transition frequency of $192\,504\,914\,418.9(4.3)$ kHz, which is in agreement with the result presented in this work and disagrees with the measurement of \cite{VanRooij2011a}.

\subsection*{Correcting the ac-Stark shift data}
For the measurements of \cite{VanRooij2011a}, the Fermi gas parameters could not be retrieved reliably, and therefore a direct correction of the data based on a re-evaluation of the spectral line shapes is not possible. However, the recorded values of the full width at half maximum (FWHM) from those measurements show a clear power dependence for $^3$He, which is absent for $^4$He, indicating the change of the Fermi-Dirac distribution as a function of the trap depth. Applying the model developed in \cite{Juzeliunas2001} to calculate expected line profiles over a wide range of Fermi gas parameters, we found that a Gaussian fit to these profiles shows a direct correlation between the peak center shift and the FWHM, independent of the specific input parameters. Hence, the recorded values of the FWHM from \cite{VanRooij2011a} can be used to correct for the nonlinearity in the ac-Stark shift. To disentangle the width contributions from the spectroscopy laser and the Fermi-Dirac distribution, we have used a fit to the measured dependence of the FWHM to the trap power using a function with two width contributions, added in quadrature: The first is a constant width related to the spectroscopy laser, the second contribution represents the thermal broadening and depends on the trap power. The measured transition frequencies were then corrected by an estimate of the thermodynamic shift, calculated through its direct correlation with the spectral line width FWHM. The corrected data and the new extrapolation are shown in Figure \ref{fig:Thermo_shift}\textbf{c}.

The linear extrapolation through this full set of corrected data resulted in an intercept at zero laser power of $192\,504\,914\,417.6(2.0)$ kHz, and a slope that fully agrees with the ac-Stark shifts measured in $^4$He. Since the $^4$He BEC mainly occupies the lowest trap state and therefore its ac-Stark shift reflects the intensity at the center of the dipole trap, this agreement between the slopes indicates that the applied correction indeed retrieves the ac-Stark shift expected from the peak intensity, further confirming that this explanation of the observed nonlinearity is valid. 

As an illustration of the influence of this systematic effect, we present a revisited value of $\delta r^2$ for the measurements of \cite{VanRooij2011a} in Figure \ref{fig:DNCR} of the main text. Originally, the measurements were split in two sets of 10 days, one set for each isotope. For the new extrapolation shown in Figure \ref{fig:Thermo_shift}\textbf{c}, we combined all measurement days into one single extrapolation in order to have sufficient statistics. To stay consistent and take into account slight variations of the dipole trap alignment over different days, we have also combined all the $^4$He measurements and performed a single ac-Stark extrapolation for these data, arriving at a value of $192\,510\,702\,144.7(2.4)$, which is within $0.5\sigma$ agreement with the originally presented result. Based on these two extrapolations, the revisited value $\delta r^2 = 1.064(15)\ \mathrm{fm^2}$ was found, as presented in Figure \ref{fig:DNCR}\textbf{c} of the main article.
\newline

\section*{Systematic Effects of the new measurement}

\subsection*{Fermi gas in a magic wavelength trap}
Under the current experimental conditions with the magic wavelength trap, the contributions to the overall shift induced by the trapping light are small, and we can develop the model of \cite{Juzeliunas2001} to find the dependence of the shift on the gas parameters. The absorption lineprofile $\mathcal{S}(\Delta)$ is given as a function of the laser detuning $\Delta$ as:
\begin{widetext}
\begin{equation}
\mathcal{S}(\Delta) \propto \int_0^\infty \mathrm{d}x\ x^2 \ln \left\{ 1+\exp\left[ \beta\mu-x^2-\dfrac{\hbar\beta}{4\omega_\mathrm{rec}}\left( \Delta + \dfrac{m_\mathrm{ex}}{\beta\hbar} x^2 \right)^2 \right]  \right\},
\label{eq:juz_general}
\end{equation}
\end{widetext}
with $m_\mathrm{ex} = 1-\bar{\omega}_e^2/\bar{\omega}_g^2$ representing the difference in trapping potential experienced by the two spectroscopy states $|g\rangle,|e\rangle = |2^3\mathrm{S}_1\rangle,|2^1\mathrm{S}_0\rangle$, $\beta = 1/k_BT$ the inverse temperature, $\mu$ the chemical potential, and $\hbar \omega_\mathrm{rec}$ the recoil energy from the absorption of a spectroscopy photon.
By performing the change of variable $\tilde{x} = x\sqrt{ 1+\dfrac{m_\mathrm{ex}}{2\omega_\mathrm{rec}}\Delta }$, it can be shown that the line profile for $m_{ex}\rightarrow 0$ can be developed in terms of the line profile at $m_{ex}=0$ to first order as:
\begin{equation} 
\mathcal{S}(\Delta)\Big|_{m_\mathrm{ex}\rightarrow 0} \underset{\sim}{\propto} \dfrac{\mathcal{S}(\Delta)\Big|_{m_\mathrm{ex}=0}}{\left( 1+\dfrac{m_\mathrm{ex}}{2\omega_\mathrm{rec}}\Delta \right)^{3/2}}.
\end{equation}
To find the shift $\Delta_{th}$ where the profile has a peak ($\Delta_{th} = 2\pi\times\Delta f_{th}$ from Figure \ref{fig:Thermo_shift}), we set its derivative to $\Delta$ equal to 0, which ultimately yields the equation
\begin{equation}
\dfrac{\partial \mathcal{S}(\Delta)}{\partial \Delta}\Bigg|_{m_\mathrm{ex}=0} = \dfrac{3}{2}\dfrac{\ \mathcal{S}(\Delta)\Big|_{m_\mathrm{ex}=0}} { \Delta + 2\omega_\mathrm{rec}/m_\mathrm{ex} } .
\label{eq:equation_delta}
\end{equation}
This equation is general and its solution gives the position of the center frequency for any Fermi gas in the limit of $m_\mathrm{ex}$ close to 0. An explicit expression can be found by considering the thermal case, for which $\beta\mu\rightarrow -\infty$, then $e^{\beta\mu}\rightarrow 0$. In this limit, one can develop $\mathcal{S}(\Delta)\Big|_{m_\mathrm{ex}=0}$ to first order and obtain a Gaussian profile:
\begin{align}
\mathcal{S}(\Delta)\Big|_{m_\mathrm{ex}=0} & \propto e^{\beta\mu} e^{-\hbar\beta\Delta^2/4\omega_\mathrm{rec}}\int_0^\infty \mathrm{d}x\ x^2 e^{-x^2}\nonumber\\
 & \propto \dfrac{\sqrt{\pi}}{4}e^{\beta\mu} e^{-\hbar\beta\Delta^2/4\omega_\mathrm{rec}}.
\end{align}
Substituting this into equation \eqref{eq:equation_delta}, we retrieve a polynomial equation for $\Delta_{th}$:
\begin{equation}
(\Delta_{th})^2 + \dfrac{2\omega_\mathrm{rec}}{m_\mathrm{ex}}(\Delta_{th}) - \dfrac{3\omega_\mathrm{rec}}{\hbar}k_B T = 0,
\end{equation}
which has generally two solutions, but one can be ignored as it diverges in the limit $m_\mathrm{ex}\rightarrow 0$. In this limit, the expression for the shift at the peak of the profile is given by the other solution:
\begin{equation}
\Delta_{th} = -\dfrac{3 m_\mathrm{ex}k_B}{2\hbar}\, T.
\end{equation}
Since it was found that the temperature in the magic wavelength trap depends linearly on the trap power, we therefore expect that a linear extrapolation of the differential ac-Stark shift, induced by a trap close to the magic wavelength, should correctly predict the unperturbed transition frequency. For the case of a degenerate gas, the states below the Fermi energy are more densely occupied, increasing their contribution. Also, due to the Pauli blockade of stimulated emission, it was shown \cite{Jannin2022} that the line profiles are symmetrically narrowed compared to the absorption profile from equation~\eqref{eq:juz_general}. Both these effects for a degenerate gas are thus expected to reduce the overall effect of this thermodynamic shift compared to the thermal case considered here.

Even if there still were a remaining nonlinear contribution to the ac-Stark shift, it is symmetric around the magic wavelength condition, and therefore cancels out in the final analysis as the measurements are symmetrically distributed around the magic wavelength. To investigate this we have performed a fit of the 11 extrapolated transition frequencies from Figure \ref{fig:transition_result}\textbf{c} of the main text, as a function of the dipole trap wavelength used for each set. This fit shows a minor wavelength dependence, but the prediction for the transition frequency at the magic wavelength is fully consistent and even more precise than the result presented in the main article. As we do not have sufficient statistics to confidently attribute this minor wavelength dependence to a systematic effect from the dipole trap, we maintain the compatible and more conservative determination shown in Figure \ref{fig:transition_result}c of the main text.

\subsection*{Vector- and tensor-polarizability}
In the magic wavelength dipole trap, the scalar polarizability for the two spectroscopy states is equal. However, the vector and tensor contributions to the polarizability are not necessarily equal since they depend on the magnetic substates that are measured, the polarization of the dipole trap light, and the direction of the quantization axis. Non-scalar contributions to the polarizability do not influence the outcome of the transition frequency since they disappear in the regression to 0 trap power. Any possible influence is limited to the determination of the magic wavelength. We estimated these non-scalar contributions to the polarizability and found that they can be neglected at our level of accuracy.
\\ \\
To estimate the tensor and vector polarizability in $^3$He we extended the calculations of \cite{Notermans2014} and used exactly the same approach used for $^4$He as described in the supplementary material of \cite{Rengelink2018}. At the magic wavelength, the vector- and tensor contributions to the polarizability of the $2^3\mathrm{S}_1(F = \frac{3}{2})$ state are, respectively:
\begin{equation}
    \alpha^{V} = 0.09 a_0^3
\end{equation}
\begin{equation}
    \alpha^{T} =  0.03 a_0^3
\end{equation}
This is the same as for $^4$He. Contrary to $^4$He $2^1\mathrm{S}_0(J = 0)$, the $2^1\mathrm{S}_0(F = \frac{1}{2})$ state in $^3$He has a vector polarizability because of the hyperfine structure. At the magic wavelength this vector polarizability is:
\begin{equation}
    \alpha^{V} =  5\times10^{-6} a_0^3
\end{equation}
These non-scalar contributions are small compared to the scalar polarizability of the spectroscopy states of $189.3a_0^3$ at the magic wavelength. Moreover, the 0.15 pm uncertainty on the determined magic wavelength represents an error in the polarizability of $0.7a_0^3$, which is also larger than any of the non-scalar contributions. Therefore these contributions to the differential polarizability can be neglected at this level of accuracy.

\subsection*{\label{methods:Collisions}Collisional Shift for a $^3$He DFG}
For bosonic $^4$He, the collisions between the atoms in the condensate result in a mean-field shift, which was measured for the first time in \cite{Rengelink2018}. In contrast, for the measurement in fermionic $^3$He the influence of collisions is negligible, with an estimated collisional shift below $1$ Hz. The most important reason is that, following from the Pauli exclusion principle, only odd partial waves contribute to the collisions between identical fermions \cite{Ketterle2007}. In the temperature range of the Fermi gases used in this work, orders of magnitude below the centrifugal barrier for $p$-wave and higher-order collisions, this means that practically the particles do not collide as long as they are indistinguishable, which also causes the requirement of sympathetic cooling during the preparation of the $^3$He degenerate Fermi gas. The absence of frequency shifts related to collisions when probing identical fermions has been shown experimentally \cite{Zwierlein2003,Gupta2003}.
During the excitation, however, the evolution between the spectroscopy states may be inhomogeneous and therefore cause dephasing between different parts of the gas, lifting the indistinguishability between the atoms. 

As an illustration, consider a pair of fermions within the gas, each of which are initially in two different motional states with the same electronic degree of freedom $|g\rangle$ (denoted here $|g_1\rangle$ and $|g_2\rangle$). The coupling induced by the spectroscopy laser to the internal state $|e\rangle$ will then transform the two states to the following superpositions:
\begin{align*}
|g_1\rangle & \rightarrow \alpha_1 |g_1\rangle + \beta_1 |e_{1}\rangle,\\
|g_2\rangle & \rightarrow \alpha_2 |g_2\rangle + \beta_2 |e_{2}\rangle,
\end{align*}
where the state index denotes the motional degree of freedom. Because of the slight inhomogeneity of evolution between the two initial states, the constructed antisymmetrized (singlet) state, which is the state in which the pair can collide according to the Pauli principle, reads \cite{Campbell2009}:
\begin{equation}
|S\rangle = \left( \alpha_1\beta_2 - \alpha_2\beta_1 \right) \left( |ge\rangle - |eg\rangle \right)/\sqrt{2},
\end{equation}
with the following probability:
\begin{equation}
\langle S | S \rangle = 1-\big| \alpha_1\alpha_2^* + \beta_1\beta_2^* \big|^2.
\label{eq:antisym_prob}
\end{equation}
It is clear that for a fully coherent excitation, where all particles evolve identically, $\alpha_1=\alpha_2$ and $\beta_1=\beta_2$, and the probability for the pair to be in the colliding $|S\rangle$ state is 0 (since $|\alpha_i|^2 + |\beta_i|^2 = 1$). However, in the extreme case where only one particle has evolved under the excitation, $\alpha_2=\beta_1=1$ and $\alpha_1=\beta_2=0$, and the pair is fully in $|S\rangle$, resulting in a collision in the \textit{s}-wave channel.

To estimate the influence of excitation inhomogeneity for the full Fermi gas, where the main cause of excitation inhomogeneity is the difference in coupling strengths between vibrational levels in the dipole trap, we follow a similar approach to \cite{Campbell2009} and calculate the average Rabi frequency and its standard deviation, $\langle\Omega\rangle$ and $\Delta\Omega$, respectively, as:

\begin{align}
    \langle\Omega\rangle &= \sum_{n} n_{g_n}(0)\ \Omega_{g_n,e_{n+l_0}}, \\
    \langle\Omega^2\rangle &= \sum_{n} n_{g_n}(0)\ \Omega_{g_n,e_{n+l_0}}^2, \\
    \Delta\Omega^2 &= \langle\Omega^2\rangle - \langle\Omega\rangle^2,
    \label{eq:avg_omega}
\end{align}    
where $n_{g_n}$ represents the Fermi-Dirac distribution and $\Omega_{g_n,e_{n+l_0}}$ is the Rabi frequency coupling the states $|g_n\rangle$ and $|e_{n+l_0}\rangle$ (see equation (70) from \cite{Leibfried2003} for details). $l_0$ represents the detuning of the spectroscopy laser expressed in terms of vibrational quanta.
The collisional shift is then calculated using the collisional energy between fully distinguishable particles in $|g\rangle$ and $|e\rangle$, modified by a two-body correlation function $G^{(2)} = \langle S | S \rangle$. $G^{(2)}$ represents the probability for two particles evolving with $\Omega_{1,2}=\langle\Omega\rangle \pm \Delta\Omega$, respectively, to become distinguishable. Using equation \eqref{eq:antisym_prob} this time-dependent correlation function can be expressed as:

\begin{equation}
G^{(2)}(t) = \sin^2 (\Delta\Omega t).
\end{equation}

\begin{figure}
    \centering
    \includegraphics[width = 0.49\textwidth]{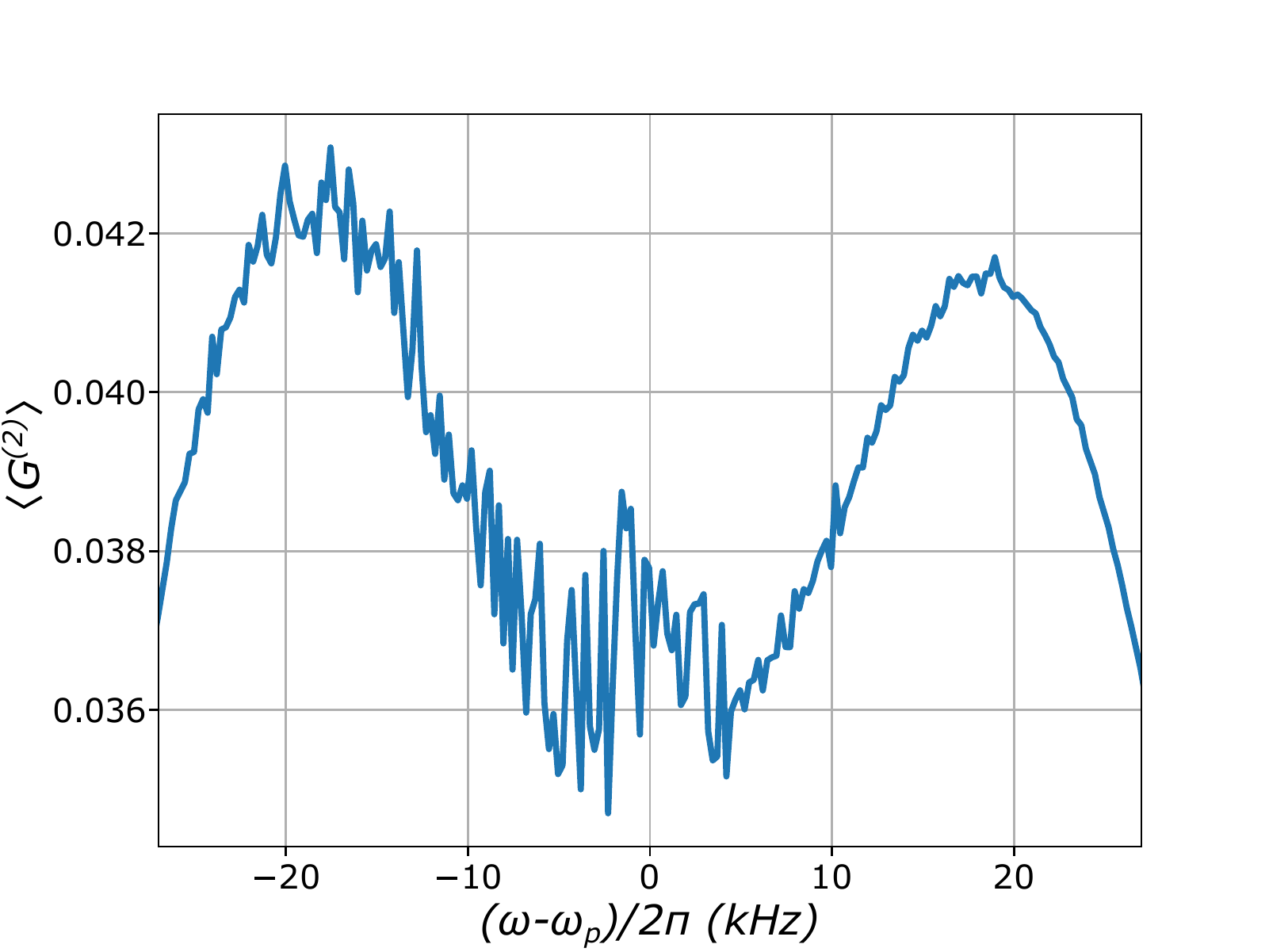}
    \caption{The time-averaged two-body correlation function $\langle G^{(2)} \rangle$ that represents the probability of \textit{s}-wave collisions due to inhomogeneous excitation, shown as a function of detuning from the peak of the spectral line profile ($\omega_p$). Parameters are: $T=0.2\ \mathrm{\mu K}$, $\mu = 0.1\ \mathrm{\mu K}$, $\omega_{ax} = 2\pi\times25$ Hz, and $\omega_{rad} = 2\pi\times 150$ Hz.}
    \label{fig:correlation_func}
\end{figure}

To find the contribution to the collisional shift requires integrating this correlation function over the time $T$ of the spectroscopy, but since the decay to the $|1^1\mathrm{S}_0\rangle$ ground-state projects the state $|e\rangle$ at a rate of $\Gamma$, we consider only the propagation over a time $1/\Gamma$:
\begin{align}
\langle G^{(2)}(t)\rangle &= \dfrac{1}{T}\int_0^{T} \mathrm{d}t\ G^{(2)}(t) \nonumber \\
&= \Gamma \int_0^{1/\Gamma} \mathrm{d}t\  \sin^2\left( \Delta \Omega t \right) \nonumber \\
&= 1- \mathrm{sinc}  \left( \dfrac{2\Delta\Omega}{\Gamma }\right). 
\end{align}
Evaluating $\Delta \Omega$ from equation \eqref{eq:avg_omega}, we can calculate the time-averaged correlation function for a range of laser detunings as shown in Figure \ref{fig:correlation_func}. At the center of the spectroscopic line profile, the value is $\langle G^{(2)} \rangle \approx 0.037$.

The collisional shift of an $s$-wave collision between an atom in the $2^3\mathrm{S}_1$ state and an atom in $2^1\mathrm{S}_0$ is calculated using the same mean-field approach that was used for $^4$He \cite{Rengelink2018}:
\begin{equation}
    \Delta E_{mf} = \frac{4\pi\hbar^2\ a_{ts}}{m}\times n_p,
\end{equation}
where $a_{ts}$ is the $s$-wave scattering length between the triplet and singlet states and $n_p$ is the density of the Fermi gas, for which we take the peak density of a fully degenerate Fermi gas as a conservative upper bound for the shift. We note that this peak density for a Fermi gas is typically over an order of magnitude lower than for a dense Bose Einstein condensate. Lacking a theoretical or experimental value for $a_{ts}$ for the $^3$He isotope, we estimate the shift using the value of $a_{ts} = 82.5(5.2)\, a_0$ ($a_0 = 0.0529$ nm being the Bohr radius), determined from the observed $^4$He mean-field shift in \cite{Rengelink2018}. This results in a total conservative upper bound for the collisional shift in a $^3$He Fermi gas of:
$$\Delta f = \dfrac{1}{h} \Delta E_{mf} \times G^{(2)} \simeq 0.02\ \mathrm{kHz} \times 0.037 < 1\ \mathrm{Hz}. $$ 
Therefore, the effect of the collisional shift that has been a significant systematic effect for measurements in a $^4$He Bose-Einstein condensate, is completely negligible for our measurements in a $^3$He Fermi gas. 

\subsection*{Further systematics}

Some other systematics that were taken into account in this work follow the same procedure as the estimations for $^4$He in \cite{Rengelink2018}. First of all the recoil shift for $^3$He can be calculated as $-h/({2m\lambda^2})=-27.276$ kHz. It is the energy shift corresponding to the recoil imposed on the atom of mass $m$ from the absorption of a spectroscopy photon with wavelength $\lambda$ ($h$ is Planck's constant), and can be calculated to a precision far beyond the accuracy of the measurements.

The dc-Stark shift is estimated conservatively with an absolute upper bound of $1\ \mathrm{mHz}$. We use the calculation in \cite{Rengelink2018} which takes the differential dc-polarizability between the $2^3\mathrm{S}_1$ and $2^1\mathrm{S}_0$ states, giving a dc-Stark shift of $\Delta\nu/E^2 \approx 6.02\times10^{-6}\ \mathrm{Hz\, V^{-2}\, m^2}$ in $^4$He. The mass-dependent difference in polarizability between the isotopes is negligible at the level of significance of these estimations, and for the dc-Stark shift the only difference is that in this work the ion-MCP, which produced the dominant dc field in \cite{Rengelink2018}, was turned off. As a conservative rough estimate, we take a field of a maximum of $10\ \mathrm{Vcm^{-1}}$, resulting in a dc Stark shift of $<1\ \mathrm{mHz}$.

We estimate $<5\ \mathrm{Hz}$ shift due to the effect of the blackbody radiation (BBR) from the room-temperature vacuum chamber. Since the BBR spectrum at room temperature is peaked at a wavelength of $\sim 10\ \mathrm{\mu m}$, we consider it as a dc contribution compared to the optical transitions from the $2^3\mathrm{S}_1$ and $2^1\mathrm{S}_0$ state, and we approximate the BBR shift using the dc-polarizability. Using the calculations of \cite{Notermans2014} we verified that the dynamic contribution to the polarizability indeed is smaller than $6\%$ of the dc component, and thereby negligible at this level of significance. The root-mean-square value of the electric field from a blackbody at T = 300\,K is $\langle E^2 \rangle \approx$ (832\,V\,m$^{-1}$)$^2$ \cite{Ludlow2015}, resulting in 4.17 \,Hz, meaning the BBR shift is $<5\ \mathrm{Hz}$.

For the quantum-interference shift, which has been argued \cite{Zheng2017} to be of relevant influence on the measurements in \cite{CancioPastor2004,CancioPastor2012}, we present the same absolute upper bound of $\Delta_{QI}\approx 2\pi\times80$ mHz as in \cite{Rengelink2018}. The available quantum paths considered in that estimation are not affected by the presence of a hyperfine structure in $^3$He, as the hyperfine structure does not introduce new couplings into the equations, and the mass-dependent and hyperfine shifts of the levels are of negligible influence on the numbers used for the estimation.

We estimate that the second-order Zeeman effect, for our applied field of at most $4$ G, leads to a maximum shift of $0.037$ Hz and $0.051$ Hz for the triplet and singlet states, respectively. In the $^4$He atom, the second-order Zeeman shift comes from the couplings of the $2^3\mathrm{S}_1$ and $ 2^1\mathrm{S}_0$ levels with the $2^3\mathrm{P}_{J}$ and $2^1\mathrm{P}_{1}$ states, respectively, which leads to very small quadratic Zeeman shifts of respectively $2.3\ \mathrm{mHzG^{-2}}$ and $3.2\ \mathrm{mHzG^{-2}}$ for the triplet and singlet states \cite{Rengelink2018}. For $^3$He, the hyperfine interaction further splits the $2^3\mathrm{S}_1$ level into the $|F=3/2\rangle$ and $|F=1/2\rangle$ state, $6.7$ GHz apart \cite{Morton2006}. However, since all measurements are performed on the spin-stretched states $|2^3\mathrm{S}_1,F=3/2,m_F=\pm3/2\rangle$, there is no coupling with the $|F=1/2 \rangle$ hyperfine level that could contribute to a higher second-order Zeeman shift, and the values of $^4$He quoted above still hold for our measurement. This was verified by calculating the Zeeman energies of these states, exchanging the roles of $J$ and $I$ in the Breit-Rabi formula \cite{Breit1931}, since $I = 1/2$ and $J = 1$ for metastable $^3$He. 

The second-order Doppler shift, which is of relevance in many precision spectroscopy experiments, is completely negligible for our ultracold quantum gases. Since we use a single-photon transition on a trapped Fermi gas, the first order Doppler shift is large, and is actually responsible for the width of the observed line profile, together with the Pauli blockade of stimulated emission \cite{Jannin2022}. In contrast, the second order Doppler shift, which is associated with relativistic time dilation of an atom moving with respect to the lab-frame, is an effect only at the level of $10^{-20}$ for our (sub-)$\mathrm{\mu K}$ temperatures.